\documentclass[
  ,draft            
  ]
  {aipproc}

\layoutstyle{6x9}

\usepackage{wrapfig}

\begin{document}

\title{Top cross section measurements at ATLAS}
\date{April 13, 2011}

\classification{14.65.Ha}
\keywords      {Top quarks}

\author{Yasuyuki Okumura on behalf of ATLAS collaboration}{
  address={Yasuyuki.Okumura@cern.ch}
}



\newcommand{\xsectt}{$\sigma_{t\bar{t}}$}
\newcommand{\ttbar}{$t\bar t$}
\newcommand{\ipb}{$\mathrm{pb}^{-1}$}
\newcommand{\mtop}{$m_{\mathrm{top}}$}
\newcommand{\pT}{\mbox{$p_{T}$}}
\newcommand{\ET}{\mbox{$E_{T}$}}
\newcommand{\MET}{$E_{T}^{miss}$}
\newcommand{\GeV}{\mathrm{~GeV}}
\newcommand{\TeV}{\mathrm{~TeV}}
\newcommand{\MTW}{$m_{T}(W)$}
\newcommand{\HT}{$H_{T}$}
\newcommand{\Zg}{$Z/\gamma^{*}$}

\begin{abstract}
Measurements of the production cross section of top-quark pairs
($t\bar t$) in $pp$ collisions at $\sqrt{s}$ = 7 TeV are presented
using 35 \ipb\ of data recorded with the ATLAS detector at the Large Hadron Collider.
Events are selected in the single lepton (electron or muon) and
dilepton topologies with multi-jets, and large missing transverse energy.
The result is
$\sigma_{t\bar t} = 180 \pm 9\,\mathrm{(stat.)} \pm15 \,\mathrm{(syst.)}
\pm 6 \,\mathrm{(lumi.)\,pb}$, which agrees with the Standard Model prediction. 

\end{abstract}

\maketitle


\section{Introduction}

A precise measurement of the top quark pair (\ttbar) production cross section
(\xsectt) allows one to test perturbative QCD (p-QCD), which predicts \xsectt\ with uncertainties at the level of 10\%. 
Furthermore, \ttbar\ production is an important background 
in searches for the Standard Model (SM) Higgs particle and 
physics beyond the SM. New physics may also 
give rise to additional \ttbar\ production mechanisms or modification
of the top quark decay channels.
Within the SM, top quarks are predicted to decay to a $W$ boson and
a $b$-quark nearly 100\% of the time, and the decay topologies are determined by
decays of the $W$ bosons.
The single-lepton and the dilepton modes,
with branching ratios of 37.9\% and 6.5\% respectively,
give rise to final states with one or two leptons (electrons or muons), missing transverse 
energy (\MET) and jets, some of which originate from $b$-quarks.   



The ATLAS detector~\cite{atlasdet} at the LHC covers nearly the entire
solid angle
around the collision point. It
consists of an inner tracking detector surrounded by a thin
superconducting solenoid, electromagnetic and hadronic calorimeters,
and an external muon spectrometer incorporating a large
superconducting toroid magnet system. 
A three-level trigger system is used to select interesting events for
recording and subsequent offline analysis.
Only data for which all subsystems described above are fully operational
in stable beam conditions are used in this analysis. 
The application of these requirements results in a data sample of $35\pm1$~pb$^{-1}$.

\section{Object definition}
The reconstruction of \ttbar\ events makes use of electrons, muons, jets, and of
missing transverse energy, which is an indicator of undetected neutrinos.
Electron candidates are defined as electromagnetic clusters 
associated to well-measured tracks.
They are required to satisfy \pT\ $> 20$~GeV and $\mathrm{|\eta_{cluster}| <}$ 2.47
\footnote{Candidates in the barrel to endcap calorimeter transition region at 1.37~$\mathrm{< |\eta_{cluster}|<}$~1.52 are excluded.}.
To suppress the background from photon conversions, 
the track must have an associated hit in the innermost pixel layer.
Muon candidates are reconstructed from track segments 
in different layers of the muon chambers, combined with charged tracks reconstructed in the inner tracking detector.
They are required to satisfy \pT\ $> 20$~GeV and $|\eta| < 2.5$.  
Both electrons and muons are required to be isolated\footnote
{
Isolation is defined as follows: 
\ET\ deposited in the calorimeter cells within $\Delta R < 0.2$ 
around the electrons is required to be less than $4\GeV$. 
\ET\ deposited in the calorimeter cells within $\Delta R < 0.3$
around the muons is required to be less than $4\GeV$, sum of \pT\ of tracks within $\Delta R < 0.3$ around muons is required to be less than $4\GeV$,
and it is required that there are no jets reconstructed with $\pT>20\GeV$ within $\Delta R<0.4$ around the muons.
}.
Jets are reconstructed with the anti-$k_t$ algorithm 
with distance parameter $\Delta R = 0.4$ from clusters of energy deposits in the calorimeters.
Jets stemming from the hadronisation of $b$-quarks are identified 
by the long lifetime of $b$-hadrons (about 1.5~ps).
A $b$-jet can be identified with the transverse impact parameter ($d_0$) of the tracks in
the jet, where $d_0$ is defined as the distance of closest approach in the transverse
plane of a track to the primary vertex.
On the base of the impact parameter significance ($d_0/\sigma_{d_0}$) of each selected track in the jet,
a probability that a jet is produced by a light quark (light-jet probability) is evaluated for each jet.
The missing transverse energy \MET\ is constructed from the vector sum of transverse momentum of
reconstructed jets, muons, and electrons, as well as calorimeter energy deposits not associated with
reconstructed objects.

\section{Analysis in single-lepton final state}
The single-lepton \ttbar\ final state is characterized by an isolated lepton with
large \pT, large \MET\ corresponding to the neutrino from the leptonic $W$
decay, two $b$-quark jets and two light jets from the hadronic $W$ decay.
The details of the event selection which uses $b$-tagging information~\cite{conf_ljet_btag} are discussed below.

Events are triggered by single electron or single muon triggers.
Leptons with \pT\ $>20\GeV$ are 
checked to be well within the plateau region of the trigger efficiency turn-on curves.
After the trigger, the event is required to contain exactly one reconstructed 
lepton.
To reject a significant fraction of the QCD multi-jet background, the \MET\ is required
to be larger than 35~GeV in the electron channel, and 20~GeV in the muon channel. 
The transverse mass of the $W$ candidate (\MTW) is required 
to be greater than 25~GeV in the electron channel, and \MET+\MTW$>60\GeV$ in the muon channel.
Finally, the event is required to have 3 or more jets with \pT$>25\GeV$ and
$|\eta |<2.5$. 
A measurement of \xsectt\ is performed
by a fit of a multivariate likelihood discriminant $D$.
The $D$ is a projective likelihood estimator built from the following four kinematic variables:
(1) $\eta$ of the selected lepton, 
(2) event aplanarity, defined as 1.5~times the smallest eigenvalue of the momentum tensor
\footnote {The momentum tensor is defined as
$M_{ij} = { \sum_{k=1}^{N_\mathrm{objects}} p_{ik} p_{jk} }/
{ \sum_{k=1}^{N_\mathrm{objects}} p_k^2 }$, 
where $p_{ik}$ is the $i$-th momentum component and $p_k$ is the modulus of the momentum 
of object $k$.} calculated with the selected jets and lepton in the events,
(3) $H_{T,3p}$, given by the transverse energy of all jets except the two leading ones
normalized to the sum of absolute values of all longitudinal momenta in the event, and
(4) $b$-tagging information $W_{JP} = -\log_{10} P_l$, where ($P_l$) is
average of the two lowest light-jet probabilities in the event computed by the $b$-tagging algorithm.
The cross section is extracted from a binned likelihood fit of $D$ to a weighted 
sum of templates corresponding to the signal and different backgrounds, simultaneously to three samples
(3-jet, 4-jet, and equal to or more than 5-jet) in the electron and muon channel separately.
The templates used in the fit are obtained from simulation for
\ttbar\ signal, $W$+jets, \Zg+jets, single-top and diboson ($WW$, $WZ$, $ZZ$) processes 
except for QCD multi-jet processes, in which the shapes are determined by data-driven techniques.
As many as possible of the systematic uncertainties are included in the fit as nuisance parameters.
These include the uncertainties from jet energy scale,
lepton detection and trigger efficiency, light-jet probability estimation, and template shapes.
Finally a maximum likelihood fit, which includes all systematic uncertainties and
bin-by-bin statistical uncertainties of the templates, is performed on the data to extract \xsectt.
%
An independent \xsectt\ extraction, which does not rely on $b$-tagging information, is performed using a 
likelihood discriminant which incorporates the charge and $\eta$ of the selected lepton, and the event aplanarity~\cite{conf_ljet_pretag}.

\section{Analysis in dilepton final state}
The dilepton \ttbar\ final state is characterized by two isolated leptons with
large \pT, large \MET\ corresponding to the two neutrinos from the leptonic $W$
decays, and two $b$-quark jets. 
The details of the event selection are discussed below~\cite{conf_dilepton}.

Events are triggered by the single-electron or single-muon triggers, and
each event is required to contain exactly two reconstructed leptons after the trigger,
which corresponds to  $ee$, $e\mu$, and $\mu\mu$ final states.
The selection of events in the dilepton channel consists of a series of
kinematic requirements on the reconstructed objects with respect to the \MET, the
invariant mass of $\mu\mu$ and $ee$ ($Z$~mass veto), the \HT\, and the number of jets reconstructed 
with $|\eta|<2.5$ and \pT$>20\GeV$. 
It is required that there are at least 2 reconstructed jets.
In order to suppress backgrounds from \Zg+jets and QCD multi-jet events
in the $ee$ and $\mu\mu$ channels, \MET\ is required to be larger than $40\GeV$, 
and the invariant mass of the two leptons must be greater than $15\GeV$.
Furthermore to reject $Z$+jets events effectively, 
the invariant mass of the two leptons is required to differ from the $Z$-boson mass
of 91~GeV by at least $10\GeV$.
For the $e\mu$ channel, the event \HT, defined as
the scalar sum of the transverse energies of the two leptons and all
selected jets, must satisfy \HT$>130\GeV$.
A measurement of \xsectt\ is performed by counting the selected events and subtracting the expected number of background events.
The \Zg\ events can be background due to mis-measurement of \MET.
The mis-measurement rate is determined by a comparison between data and simulation in \Zg+jets control samples.
The QCD and $W$+jets events can mimic signal candidates if one or two jets are mis-identified as isolated leptons.
The fake rate of lepton identification is measured, and the selection rate for these backgrounds 
are determined according to the measured fake rates.
Other SM background rates are estimated with predictions from simulation.
An independent \xsectt\ extraction with $b$-tagging techniques is performed with the additional requirement for events to include at least one $b$-tagged jet, and
loosened kinematic selections.

\section{Results of cross section measurements}

\newcommand{\xsecdileptontot}{173}
\newcommand{\xsecdileptonstat}{\pm 22}
\newcommand{\xsecdileptonsyst}{^{+18}_{-16}}
\newcommand{\xsecdileptonlumi}{^{+8}_{-7}}
\newcommand{\xsecdileptonbtot}{171}
\newcommand{\xsecdileptonbstat}{\pm 22}
\newcommand{\xsecdileptonbsyst}{^{+21}_{-16}}
\newcommand{\xsecdileptonblumi}{^{+7}_{-6}}


\newcommand{\ttbareeNJetsTwoJet}{11.5 $\pm$ 1.3}
\newcommand{\ttbarmmNJetsTwoJet}{$20.1\pm 1.7$}
\newcommand{\ttbaremNJetsTwoJet}{$47.4 \pm 4.0$}
\newcommand{\DataeeNJetsTwoJet}{16}
\newcommand{\DatammNJetsTwoJet}{31}
\newcommand{\DataemNJetsTwoJet}{58}
\newcommand{\TotalNonttbaree}{3.5 $\pm$ 1.1}
\newcommand{\TotalNonttbarmm}{$7.3^{ +1.8}_{ -1.5}$}
\newcommand{\TotalNonttbarem}{$10.8 \pm 3.4 $}
\newcommand{\TotalExpectedee}{15.0 $\pm$ 1.7}
\newcommand{\TotalExpectedmm}{$27.4 \pm 2.4$}
\newcommand{\TotalExpectedem}{$58.2 \pm 5.2$}



\begin{figure}
  \label{fig:results}
  \includegraphics[height=.225\textheight]{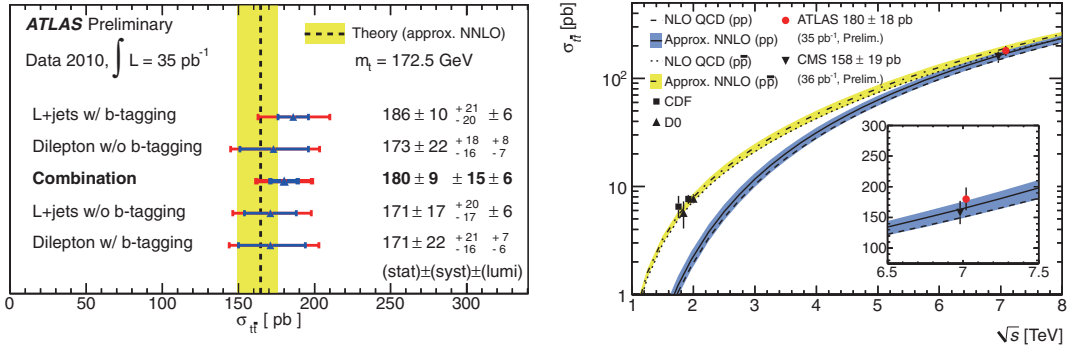}
  \caption{Left: Summary table for all the individual \xsectt\ measurements. All results are in good agreement. Right: The combined result of all the measurements is described by red closed circle. The measurement is consistent within uncertainties, with the p-QCD prediction (NNLO calculation). It is also in good agreement with the results measured at the CMS~\cite{CMS2011comb}. }
\end{figure}

\newcommand{\xseccombtot}{180}
\newcommand{\xseccombstat}{\pm 9}
\newcommand{\xseccombsyst}{\pm 15}
\newcommand{\xseccomblumi}{\pm 6}
Results of all individual measurements are summarized in Figure~\ref{fig:results}. 
The results in the single-lepton and dilepton final states are in good agreement with each other, as well as with the SM prediction.
The combined result of the single-lepton channel and the dilepton channel is
$\xseccombtot \xseccombstat (\mathrm{stat.})\xseccombsyst (\mathrm{syst.}) \xseccomblumi (\mathrm{lumi.})~\mathrm{pb}$,
which is based on the estimation of a profile likelihood ratio, taking the correlation of uncertainties between channels into account~\cite{conf_combination}.
The results using $b$-tagging information in the single-lepton channel, and results without $b$-tagging information in the dilepton channel
are combined, where the analyses which are combined are chosen to minimize the uncertainty of the measurement. 
The measurement result is in good agreement with the SM prediction as shown in Figure~\ref{fig:results}.

\begin{thebibliography}{99}



\bibitem{atlasdet}
ATLAS Collaboration,
{\it The ATLAS Experiment at the CERN Large Hadron Collider}, 
JINST 3 S08003 (2008).

\bibitem{conf_ljet_btag}
ATLAS Collaboration, 
{\it Measurement of the top quark pair cross-section with ATLAS in $pp$ collisions at $sqrt{s}$=7~TeV in the single-lepton channel using b-tagging},
ATLAS-CONF-2011-036 http://cdsweb.cern.ch/record/1337785.

\bibitem{conf_ljet_pretag}
ATLAS Collaboration, 
{\it Top Quark Pair Production Cross-section Measurement in ATLAS in the Single-Lepton+Jets Channel without b-tagging},
ATLAS-CONF-2011-024 http://cdsweb.cern.ch/record/1336753.

\bibitem{conf_dilepton}
ATLAS Collaboration, 
{\it Measurement of the top quark pair production cross section with ATLAS in $pp$ collisions at $sqrt(s)$=7~TeV in dilepton final states},
ATLAS-CONF-2011-035 http://cdsweb.cern.ch/record/1337784.

\bibitem{conf_combination}
ATLAS Collaboration, 
{\it A combined measurement of the top quark pair production cross-section using dilepton and single-lepton final states},
ATLAS-CONF-2011-041 http://cdsweb.cern.ch/record/1338569.

\bibitem{CMS2011comb}
CMS Collaboration
{\it Combination of top pair production cross sections in pp collisions at $\sqrt{s}$ = 7 TeV and comparisons with theory}, CMS-PAS-TOP-11-001 
http://cdsweb.cern.ch/record/1336491.






\end{thebibliography}
\end{document}